# Закон сохранения энергии в гидродинамике vs понятие альтернативной энергии. Критические заметки по поводу статьи «Альтернативная энергетика vs лженаука» и цитированных и нецитированных в ней работ.

*Igor Sokolov, University of Michigan, 2455 Hayward Str., Ann Arbor MI 48109*
*e-mail: igorsok@umich.edu*

## 1. Введение

*Здесь дан критический обзор серии публикаций в АЭЭ по «бесплотинным электростанциям» [1-10]. Сопоставление с общеизвестным уравнением энергии в гидродинамике демонстрирует удручающе низкий уровень всех без исключения работ (как за, так и против), а также и рецензий в данной серии.*

Начнем с [9-10], где утверждается, что работы [2-5], ранее опубликованные в АЭЭ и описывающие двухступенчатую «бесплотинную электростанцию», лженаучны и противоречат закону сохранения энергии: невозможно, чтобы после такой турбины поток воды ускорялся, уровень потока понижался и энергии вырабатывалось гораздо больше поступающей в турбину кинетической энергии. Отметим «заказной» характер рецензий [6-10] написанных "После того, как был официально поднят вопрос о ложности идеи двухколесной гидротурбины…" [6]. Само по себе это несущественно: и экспертизы, и рецензии – это заказные материалы. Но вот нецитирование в этом контексте статьи [1], также опубликованной в журнале АЭЭ, как минимум, тенденциозно, тем более, что в [2-5] эта работа цитируется и многие ее утверждения воспроизводятся. А между тем в [1] будто бы экспериментально показано, что за «бесплотинной электростанцией» поток ускоряется с 1 м/с до 3 м/с, уровень потока в следе за турбиной значительно понижается и электроэнергии вырабатывается чуть ли не в десять раз больше, чем причитается по закону сохранения энергии. При этом если уж редакция журнала АЭЭ публиковала в [1] эти непонятно кем и как как померенные цифры и непонятно откуда взятые утверждения как установленный экспериментальный факт, то как-то странно публиковать восемь лет спустя рецензии о том, что (в общем-то в аналогичной схеме) достичь такого результата в принципе невозможно. Таким образом ответ на вопрос, научна ли идея «бесплотинной электростанции» или нет, необходимо включает анализ научного содержания исходной работы [1]. К тому же [1-10] основаны на странном представлении о законе сохранения энергии в гидродинамике, с изложения которого мы и начнем.

## 2. Закон сохранения энергии в гидродинамике.

Закон сохранения энергии сжимаемой жидкости (силой тяжести пока пренебрегаем) описывается уравнением (6.1) из [11]:

$$\frac{\partial}{\partial t}\left(\rho \frac{u^2}{2} + e\right) + \nabla \cdot \left[\vec{u}\left(\rho \frac{u^2}{2} + e\right) + \vec{u}P\right] = 0 \quad , \quad (1)$$

где $\rho, \vec{u}, e, P$ - массовая плотность, скорость, плотность внутренней энергии и давление соответственно. Рассматривается одномерное течение в прямоугольного сечения канале шириной L. Проинтегрируем (1) по объему, ограниченному двумя поперечными сечения канала, 1 и 2 (поток направлен от 1 к 2):

$$\frac{d}{dt}\int\left(\rho\frac{u^2}{2}+e\right)dV = \int \vec{u}\left(\rho\frac{u^2}{2}+e+P\right)\cdot d\vec{S}_1 - \int \vec{u}\left(\rho\frac{u^2}{2}+e+P\right)\cdot d\vec{S}_2 \quad ,$$

т.е. скорость приращения полной энергии (кинетической и внутренней) равна разности между втекающим в этот объем потоком энергии через сечение 1 и потоком энергии, вытекающим через сечение 2. Поток энергии включает работу сил давления $\int P\vec{u}\cdot d\vec{S}$ и адвекцию удельной (на единицу массы) полной энергии, $\frac{u^2}{2}+\frac{e}{\rho}$, последнее видно из сопоставления с законом сохранения массы:

$$\frac{\partial \rho}{\partial t}+\nabla\cdot(\rho\vec{u})=0, \qquad \frac{d}{dt}\int \rho dV = \int \rho\vec{u}\cdot d\vec{S}_1 - \int \rho\vec{u}\cdot d\vec{S}_2 . \qquad (2)$$

В приложении к «бесплотинным гидроэлектростанциям» учтем силу тяжести и, воспользуемся приближением несжимаемой жидкости, (см. [11]):

$$\rho = \text{const}, \quad \nabla\cdot\vec{u}=0, \quad \rho\left[\frac{\partial \vec{u}}{\partial t}+(\vec{u}\cdot\nabla)\vec{u}\right]+\nabla P + \rho\nabla(gh)=0, \qquad (3)$$

где h - вертикальная координата, $q = 9{,}8$ м/с$^2$ – ускорение силы тяжести. Умножим третье уравнение в (3) на вектор скорости и прибавим к нему первое из уравнений (2), умноженное на $\frac{u^2}{2}+gh$. После тождественных преобразований с учетом второго из уравнений (3) и равенства $\frac{\partial(gh)}{\partial t}=0$ получим уравнение, аналогичное (1) и выражающее закон сохранения энергии в несжимаемой жидкости (в архаичных обозначениях данное уравнение приведено в [15]):

$$\frac{\partial}{\partial t}\left[\rho\left(\frac{u^2}{2}+gh\right)\right]+\nabla\cdot\left[\rho\vec{u}\left(\frac{u^2}{2}+gh\right)+\vec{u}P\right]=0$$

В плотность энергии включена потенциальная энергия поля силы тяжести. В поток энергии вносят вклад: поток кинетической энергии, поток потенциальной энергии и работа сил давления. Разность потоков энергии через два поперечных сечения канала, 1 и 2, в нестационарных (например, таких, как распространении волн по каналу) процессах может быть равна росту энергии жидкости в объеме, ограниченном сечениями 1,2. Однако в обсуждаемых приложениях нас интересует случай, когда в установившемся течении (d/dt=0) внутри этого объема «прирастает» не энергия потока, а производимая за счет энергии потока «негидродинамическая» энергия, как результат производимой электрической мощности, E, и диссипируемой во всех источниках потерь мощности W:

$$E+W = \int\left[\rho\vec{u}\left(\frac{u^2}{2}+gh\right)+\vec{u}P\right]\cdot d\vec{S}_1 - \int\left[\rho\vec{u}\left(\frac{u^2}{2}+gh\right)+\vec{u}P\right]\cdot d\vec{S}_2 \qquad (4)$$

(математические изыски отнесены в Приложение).

Вычислим интегралы: предполагая, что сечения 1 и 2 выбраны достаточно далеко против- и по- потоку от «бесплотинной электростанции», так что течение в сечениях можно считать однородными; пренебрегая перепадом уровней дна; учитывая сохранение массы (интегралы в правой части в (2) равны); считая давление равным нулю на свободной поверхности и нарастающим вглубь по закону Паскаля: $P_{1,2} = \rho g(H_{1,2}-h)$, $H_{1,2}$ – высота свободной поверхности. Имеем:

$$H_1 u_1 = H_2 u_2, \qquad \frac{E+W}{L\rho H_1 u_1} = \left(gH_1+\frac{u_1^2}{2}\right)-\left(gH_2+\frac{u_2^2}{2}\right). \qquad (5)$$

При фиксированных входных потоках массы и энергии производимая энергия и выходящий поток энергии зависят от конструкции «электростанции»: чем больше производимая энергия и потери, тем меньше выходящий поток энергии. Последний ограничен снизу значением $\min\left(gH_2+\frac{u_2^2}{2}\right)=\frac{3}{2}gH_1(Fr_1)^{1/3}$, достижимым при $H_2 = H_1(Fr_1)^{1/3}$, $u_2 = u_1(Fr_1)^{-1/3}$. Здесь $Fr_1 = \frac{u_1^2}{gH_1}$ – число

Фруда во входящем потоке. В «идеальной конструкции» электростанции, в которой достигается максимум производимой энергии, в зависимости от числа Фруда поток либо ускоряется (при малом числе Фруда $Fr_1 < 1$, например, в водохранилище с высоким уровнем и медленной скоростью потока перед плотиной, ускорение потока сопровождается значительным понижением уровня за плотиной), либо тормозится на гидродинамическом сопротивлении турбины (при высоком числе Фруда $Fr_1 > 1$), на выходе «идеальной конструкции» удельная энтальпия минимальна, $gH_2 + \frac{u_2^2}{2} = \frac{3}{2}gH_1(Fr_1)^{1/3}$, и число Фруда равно единице: $Fr_2 = \frac{u_2^2}{gH_2} = 1$. При этом выходящий поток энергии может быть значительно меньше входящего, что означает эффективное преобразование энергии:

$$E + W = \rho L u_1 g H_1^2 (1 + \frac{Fr}{2} - \frac{3}{2} Fr^{\frac{1}{3}}) \tag{6}$$

Наоборот, при малом отличии выходного потока энергии от входного (при $H_1 \approx H_2$, энергетическая эффективность «электростанции» пренебрежимо мала.

### 3. Работы [2-10].

С учетом изложенных прописных истин учебного характера, рассмотрим, начиная с последней, все работы из серии [1-10]. В работах [9-10] (то же самое в [6]) в потоке энергии упущен член с давлением, что приводит к тройной ошибке: 1.) публикации [2-5] необоснованно критикуются за неучет работы, которая должна быть совершена для изменения кинетической энергии потока (как раз в [2-5] используются формулы, учитывающие вклад давления, и именно силы давления в сплошной среде совершают работу) 2.) критикуются уравнения (5-6) и 3.) заявляется, что «критическое» значение числа Фруда, при котором отбор энергии от потока должен сопровождаться ускорением потока, равно не 1 а ½. Рецензии [7,8] учитывают только вклад от кинетической энергии в поток энергии и оценивают максимально возможное производство энергии в потоке со скоростью 1 м/сек и сечением канала 1 м² как 0,5 кВт= $\rho L H_1 u_1^3/2$. Такой подход вызывает недоумение, тем более что запретительного характера численные оценки сделаны для параметров потока на стенде (см далее), предназначенного для испытания моделей с $Fr_1 < 1$, для которых вклад кинетической энергии в поток энергии перед устройством мал по сравнению с давлением и потенциальной энергией. Должен заметить, что после исправления этих очевидных ошибок от материалов [7-10], на мой взгляд, ничего не остается – кроме приводящего в изумление тона и стиля.

Далее, работы [2-5] безусловно и элементарно неверны, и их публикацию следовало бы считать (четырежды повторившейся) ошибкой со стороны редакции журнала АЭЭ. В этих работах в качестве оптимальной «бесплотинной электростанции» в потоке с небольшим (~0,1) числом Фруда предлагается (неважно, как) реализовать следующую схему преобразования потока: за «бесплотинной электростанцией» поток значительно ускоряется и, соответственно, уровень воды непосредственно за сооружением значительно понижается и затем несколько ниже по потоку возникает «гидравлический прыжок», повышающий уровень воды в уходящем потоке почти до того же значения, что имеет входящий поток. Отсюда, располагая сечение 2 ниже по потоку, чем гидравлический прыжок (соответственно, для предложенной схемы преобразования потока имеем $H_2 \approx H_1$), получаем, что сумма E+W произведенной мощности и мощности диссипативных потерь мала в силу (5) при $H_2 \approx H_1$. Но мало того, что сумма W+E мала, но еще и W в такой схеме велико [12]! Перепад высот в гидравлическом прыжке возникает за счет вихреобразования, которое влечет за собой значительную турбулентную диссипацию в области прыжка. Малость суммы W+E в совокупности с ростом W заставляет сомневаться, будет ли в такой схеме положительное производство энергии E>0, или наоборот, чтобы реализовать такую схему, должна тратиться энергия (E<0) на ускорение реки.

К достоинствам работ [2-5], можно отнести использование формул для оценки энергетической эффективности «идеальной конструкции», и то, что по потоку за «идеальной конструкцией» действительно возможно отмеченное в [2-5] образование стоячих возмущений (например, таких как гидравлический прыжок), поскольку за «идеальной конструкцией» скорость длинных волн равна скорости потока, $u_2 = \sqrt{gH_2}$, и волны не сносятся ни по потоку, ни против. Недостаток работ [2-5], по-видимому, неустранимый и справедливо отмеченный в [9-10]: нет доказательств, что предложенная конкретная схема «двухступенчатой бесплотинной электростанции» является шагом в направлении «идеальной конструкции» или хотя бы воспроизводит обсчитываемую конфигурацию течения.

**5. Работа [1], часть 1**

Публикация [1] состоит из двух частей. В первой части описано устройство, дважды названы его размеры: 1,25 м (длина) на 1,2 м (ширина) на 0,7 м (высота). При этом на фото изображено заведомо не такое устройство и во второй части работы другим Автором приведены другие размеры 1,5×0,6×0,6 м.

Для устройства с заявленными размерами и конструкцией голословно утверждается возможность генерации 315 кВт электрической мощности при погружении в ручей со скоростью потока 10 м/с. Изображенное устройство на фото имеет не такие размеры, поток не имеет такой скорости. Значение 315 кВт получено не экспериментально, а в результате ряда абсурдных манипуляций, в качестве последней манипуляции 315,000 Ньютонов преобразованы в 315 киловатт, что одно уже означает, что эти выкладки никакого смысла не имеют. Установленным физическим фактом возможность генерации 315 киловатт электрической мощности не является. Вообще в первой части публикации [1] не установлен факт генерации какой бы то ни было электрической мощности.

В качестве подписи под рисунком дано: скорость потока перед устройством равна 1 м/с, скорость потока за устройством 3,14 м/с. Непонятно откуда взяты эти значения и как они померены. Про фото "Поток на выходе установки, скорость 3,14 м/с" скажем осторожно: в потоке с волновой турбулентностью на свободной поверхности есть много соблазнительных для дилетанта возможностей померять высокую скорость (скажем проследить на киносъемке перемещение за какого-нибудь "барашка"), но нужно быть профессионалом в гидродинамике, чтобы знать, что к скорости потока скорость такого "барашка" никакого отношения не имеет. Покуда не описан метод измерения, утверждение про трехкратное повышение скорости потока является неосновательным личным мнением Автора.

Больше никакой научной информации в первой части [1] нет, есть ссылка на авторитеты. Ознакомимся с отзывом одного из них [13]: «*На ДМЗ "Камов" в Московской области были изготовлены десять ~~движетелей~~ турбин по чертежам Ленёва Н.И. и при его авторском надзоре*» - как заявлялось в [1], их проектная мощность составляла 10 кВт, там же (во второй части) сообщалось об успешной генерации постоянного тока с мощностью 3,2 кВт. Продолжение цитаты: «*К сожалению, испытания, проведенные 13.02.06г. фирмой "ИНСЭТ" г. Санкт-Петербург с привлечением специалистов из Санкт-Петербургского Политехнического Университета, показали полную бесперспективность данных ~~движителей~~ турбин*». Акт испытаний представлен в [13]. Пока электрическая цепь электрогенератора не была включена, турбина раскручивала его до необходимой частоты, но при включении на электрическую нагрузку частота вращения генератора упала ниже рабочей частоты и генерация сорвалась. Отметим, что с разомкнутым генератором одна лишь мощность потерь, $W = 400$ Вт, уже превышает придуманный в [7,8] "теоретический предел" $\frac{\rho L H_1 u_1^3}{2} \approx 300$ Вт. Здесь, как и в случае дискуссии между авторами [2-6,9-10], заведомая неправота одной из сторон вовсе не означает правоту оппонентов.

Итак, первая часть работы [1] не содержит никакой научной информации и установленных научных фактов. Приведенные параметры устройства не соответствуют приведенным фотографиям, оценки производства электрической энергии не соответствуют фактическим параметрам устройства и потока и неверны даже по размерности, а приведенные данные по скоростям потока не сопровождаются описанием измерительной методики и достоверность их не установлена. Журнал АЭЭ ввел публикацию [1] в научный оборот под видом научного факта, в то же время о последовавшем провальном испытании устройства и прекращении (заявленного в статье!) производства турбины не сообщалось.

### 5. Работа [1], часть 2

Начало заглавия второй части "Отзыв на изобретение…", первой части "Бесплотинные ГЭС на основе...". Пусть не плотина, но некоторая преграда в течении, повышающая уровень водного потока на фото с подписью "перед загрузкой" присутствует. Это непроницаемое основание устройства от нижней кромки до верхнего края нижней направляющей, высота этой плотины приблизительно 10 см и на те же 10 см она потенциально способна поднять поток перед устройством. Много это или мало? При спуске одного килограмма воды с высоты 10 см=0,1м он теряет потенциальную энергию 0,1м х 1 кг х 9,8 м/с²=0,98 Дж. При скорости потока 1 м/с через каждый 1 м² сечения канала проходит 1000 кг/м³ х 1 м/с х 1 м² = 1000 кг/с и в случае подъема потока перед турбиной на высоту 10 см, при последующем спуске с этой высоты мощность 1000 кг/с х 0,98 Дж/кг = 980 Вт может быть использована для выработки электроэнергии. Тем самым, если на установке [1] генерируется какая-то мощность, то первый киловатт этой мощности, на мой взгляд, не только не представляет загадки, но даже не делает устройство изобретением с отличительным признаком "бесплотинная". К тому же важная информация [1] о том, что за устройством уровень потока понижается на $H_2 = -0,2$ м, оказывается неполной без сообщения об уровне перед устройством.

Узловым моментом второй части работы [1] является кажущееся расхождение между экспериментом по генерации электрической мощности ~3 кВт и оценочным расчетом. Примем пока на веру результат эксперимента. Сравним с законом сохранения энергии (5), в котором подставим следующие величины: а) расход $\rho L H_1 u_1 = 1000$ кг/с (как и принято в [1]) б) повышение уровня потока перед устройством $H_1 = 0,1$ м (оно не обсуждается в [1], но это не значит, что его нет),

в)понижение уровня потока за устройством $H_2 = -0,2$ м и г)пренебрегаем ускорением жидкости: $u_2 \approx u_1$, точнее, $|u_1^2 - u_2^2| \ll 2g|H_1 - H_2|$ . Имеем: $E + W = 1000 кг/с \times (9,8 \text{ м/с}^2 \times 0,1 \text{ м} - 9,8 \text{ м/с}^2 \times (-0,2 \text{ м})) \approx 3$ кВт . Резкого расхождения между экспериментом и законом сохранения энергии не видно.

Допустимо ли игнорирование "экспериментальных данных" по ускорению потока до 3 метров в секунду? Не только допустимо, но обязательно: подставив в (5) $u_1 = 1 м/с$ и $u_2 = 3 м/с$ , мы бы получили: $E + W = 1000 кг/с \times (9,8 \text{ м/с}^2 \times 0,1 \text{ м} + 0,5 \text{ м}^2/\text{с}^2 + 9,8 \text{ м/с}^2 \times 0,2 \text{ м} - 4,5 \text{ м}^2/\text{с}^2 ) = -1$ кВт . Такое ускорение потока противоречит закону сохранения энергии: для его реализации требуется иметь внешний источник энергии с мощностью 1 кВт. Утверждение о трехкратном ускорении потока недостоверно и вовлекаться в научный оборот не должно.

Наконец, в реальности есть некоторые потери и $W > 0$ , так что потенциальная возможность преобразования E+W=3 кВт энергии потока для генерации E=3 кВт электрической мощности не вполне достаточна. Но в реальности, описанное в [1] устройство значимой электрической мощности пока что и не генерирует, как показали испытания, и закон сохранения энергии восторжествовал. А утверждение о генерации 3 кВт электрической мощности оставим на совести журнала АЭЭ. Ни физику, ни энергетику оно не ниспровергает, разве что может наносить людям, наивно склонным доверять научным журналам, репутационный и финансовый урон [13].

**6. Заключение.**

Журнал, освещающий в том числе и проблемы гидроэнергетики, 8 лет пребывает в перманентном конфликте с законом сохранения энергии в гидродинамике. Особенно стыдно наблюдать такое в журнале, издающемся в г. Сарове: среди многих знаменитых изделий с маркой «сделано во ВНИИЭФ» есть и книга [14] и закон сохранения энергии в ней изложен не хуже чем в [11].

Благодарю за полезные замечания, обсуждения и помощь С.Д.Захарова и В.Б.Морозова (ФИАН им.П.Н.Лебедева), Д.Б.Зотьева (Волжский филиал МЭИ), А.И.Петрова (МГТУ им. Н.В.Баумана) и Г.В.Трещалова (ERG).

**7. Примечание.**

При первой переработке были добавлены пояснения при выводе закона сохранения энергии в несжимаемой жидкости (ненумерованное уравнение между (3) и (4)), добавлена ссылка [15], а также исправлены опечатки в уравнении (6) (был пропущен коэффициент 3/2 перед последним членом в скобках) и в измеренном значении величины $H_2$ (0,2 м вместо ошибочного 0,2 см). В третьем варианте добавлен (не необходимый) математический вывод уравнения (4) (см нижеследующее Приложение). В четвертом варианте исключена аннотация и добавлена дискуссионная работа «К дискуссии TZS-1-1», представленная для публикации в Письма в журнал «Альтернативная энергетика и экология».

**8. Приложение.**

Выход первого варианта настоящей работы выявил, к моему удивлению, что уравнения (4-5) рассматриваются как предмет идейной борьбы не только некоторыми авторами журнала АЭЭ, но и редакцией. В связи с этим, хотя соотношения (4-5) выражают собой несомненный факт – разность входящего и выходящего потоков энергии равна сумме мощности производимой энергии плюс мощности всех потерь – дадим чуть более подробный (и явно избыточный)

математический вывод соотношения (4) для следующего обобщения уравнения, предшествующего формуле (4):

$$\frac{\partial}{\partial t}\left[\rho\left(\frac{u^2}{2}+gh\right)\right]+\nabla\cdot\left[\rho\vec{u}\left(\frac{u^2}{2}+gh\right)+\vec{u}P\right]=-\eta, \qquad (7)$$

где в правой части добавлена мощность вязкой диссипации гидродинамической энергии на единицу объёма (подынтегральное выражение в уравнении (16.3) в [11]). Стандартным в теоретической физике (см, например, [16]) является взгляд на левую часть уравнения (7) как на четырехмерную дивергенцию, $\left(\frac{\partial}{\partial t},\nabla\right)\cdot\left[\rho\left(\frac{u^2}{2}+gh\right)(1,\vec{u})+P(0,\vec{u})\right]$ во времени-пространстве $(t,\vec{x})$ так что переход к интегральной форме уравнения (7) осуществляется его интегрированием по четырехмерному объёму, ограниченному гиперповерхностями $t=t_0$ и $t=t_0+dt$, а также сечениями канала 1 и 2 против- и по- потоку от турбины. При применении теоремы Остроградского-Гаусса в 4-х мерном пространстве, левая часть (7) преобразуется в сумму интегралов по 3-х мерным гиперповерхностям, в число которых входит, во-первых, разность интегралов энергии $t=t_0+dt$ и $t=t_0$, равная нулю для установившегося режима вращения турбины: $\int\rho\left(\frac{u^2}{2}+gh\right)dV\left|_{t=t_0}^{t=t_0+dt}\right.=0$. Кроме того в сумму входят интегралы $\int\left[\rho\vec{u}\left(\frac{u^2}{2}+gh\right)+P\vec{u}\right]\cdot\vec{dS}\,dt$ по сечениям 1 и 2, а также интегралы по движущимся с локальной скоростью $\vec{U}$ поверхностям лопаток турбин, выражение для интеграла потока энергии через которые отличается от выражения для неподвижной поверхности и равно $\int\left[\rho(\vec{u}-\vec{U})\left(\frac{u^2}{2}+gh\right)+P\vec{u}\right]\cdot\vec{dS}dt$, где вектор $\vec{dS}$ направлен по нормали к поверхности в сторону от жидкости к лопатке (при выводе последнего выражения учитывается зависимость направляющих косинусов вектора элемента гиперповерхности от скорости движения $\vec{U}$). Условие непротекания на поверхности движущейся лопатки, $(\vec{u}-\vec{U})\cdot\vec{dS}=0$, преобразует этот интеграл как $\int P\vec{u}\cdot\vec{dS}\,dt=\int P\vec{U}\cdot\vec{dS}\,dt$, причем сумма интегралов $\Sigma\int P\vec{U}\cdot\vec{dS}$ по всем лопаткам равна работе dA/dt, совершаемой потоком воды над турбиной за секунду (интегралом от скалярного произведения элементарной силы, $P\vec{dS}$ на локальную скорость). Опустим перечисление остальных интегралов по гиперповерхностям в левой части проинтегрированного уравнения (7), и объяснения, почему они равны нулю. В правой же части проинтегрированного уравнения мы получим умноженное на dt и взятое с обратным знаком выражение для мощности диссипативных потерь в жидкости: $-dtW_d=-dt\int\eta dV$, так что интегральная форма уравнения имеет вид:

$$\left\{\int\left[\rho\vec{u}\left(\frac{u^2}{2}+gh\right)+P\vec{u}\right]\cdot\vec{dS_2}-\int\left[\rho\vec{u}\left(\frac{u^2}{2}+gh\right)+P\vec{u}\right]\cdot\vec{dS_1}+\frac{dA}{dt}+W_d\right\}dt=0. \qquad (8)$$

Разделив на $dt$, получаем, что разность входящего и выходящего потоков энергии равна сумме интегральной мощности вязких диссипативных потерь и работы, совершаемой потоком воды над турбиной за единицу времени. Представляя последнюю работу в виде суммы мощности E производимой в турбине энергии и мощности $W_e$ потерь в турбине: $\frac{dA}{dt}=E+W_e$, приходим к формуле (4) с суммарной мощностью потерь, $W=W_e+W_d$.

## Список литературы.

# К дискуссии Трещалов-Зотьев-Соколов (TZS-1-1)
*Igor Sokolov, University of Michigan, USA. e-mail: igorsok@umich.edu*

*Ранее в [1] была критически проанализирована опубликованная в журнале АЭЭ серия из 10 работ, в том числе [2-8]. При этом оказалось, что в той или иной степени все рассмотренные работы не вполне согласуются с общеизвестным уравнением энергии в гидродинамике. В настоящей дискуссионной работе я обращаю внимание на то, что журнал АЭЭ несколько раз опубликовал статьи, ниспровергающие основы физики, такие как интеграл Бернулли, закон сохранения энергии в гидродинамике, волновая функция фотона и др, причем все эти основы заведомо лежат за пределами профильной тематики журнала АЭЭ. При этом кого-то из читателей в большей степени удивит утверждение о волновой функции фотона, кого-то о законе сохранения энергии - в теоретической форме или в принятой в гидравлике форме интеграла Бернулли - но совокупность фактов, на мой взгляд, не оставит равнодушным любого.*

Мой интерес к публикациям в АЭЭ случайный, но профессиональный: я обратил внимание на сведения об эксперименте [2] с неожиданными результатами по энергетике и по турбулентности, кажущимися близкими по физике к процессам нагрева и ускорения плазмы в турбулентном солнечном ветре, которыми я занимаюсь [9]. Проанализировав работы [2-8], я выяснил, что анализ и без того непростой проблемы осложнен не всегда правильным учетом закона сохранения энергии в течении воды со свободной поверхностью [1]:

$$\frac{\partial}{\partial t}\left[\rho\left(\frac{u^2}{2}+gh\right)\right]+\nabla\cdot\left[\rho\vec{u}\left(\frac{u^2}{2}+gh\right)+\vec{u}P\right]=0. \qquad (1)$$

В плотность энергии включена потенциальная энергия поля силы тяжести. В поток энергии вносят вклад: поток кинетической энергии, поток потенциальной энергии и работа сил давления. Разность потоков энергии через два поперечных сечения канала, 1 и 2, выше и ниже турбины по течению вычислим, считая в этих сечениях горизонтальное течение однородным и давление нарастающим вглубь по закону Паскаля: $P_{1,2}=\rho g(H_{1,2}-h)$, что дает [1]:

$$H_1 u_1 = H_2 u_2, \qquad \frac{E+W}{L\rho H_1 u_1} = \left(gH_1+\frac{u_1^2}{2}\right)-\left(gH_2+\frac{u_2^2}{2}\right), \qquad (2)$$

где $H_{1,2}$ – высота свободной поверхности, $L$ – ширина канала, $E$ – мощность производимой электроэнергии, $W$ – мощность диссипативных потерь. При этом в работах [3-6] неправильно, на мой взгляд, оценен вклад «гидравлического прыжка» в диссипацию энергии [1]. В статьях же [7,8] вообще не учтен вклад от работы сил давления в поток энергии. Попытки обратить внимание их автора на ошибку выявили, что по каким-то причинам он склонен опровергать соотношение (2) (одну из форм закона сохранения энергии) многими способами, разумеется, ошибочными - рецензии с анализом неудачных попыток такого опровержения представлены в редакцию АЭЭ. По моему мнению, работы [7,8], которые необоснованно объявляют статьи [3-6] лженаучными и противоречащими закону сохранения энергии, несостоятельны и не имеют веса**,** поскольку сами основаны на ошибочных представлениях о формулировке этого закона в гидродинамике. Вообще признание какого-то направления работ лженаукой (см заглавие статьи [8], прошедшей рецензирование и экспертизу) требует гораздо более высокого уровня ответственности и компетентности, чем тот, который, на мой взгляд, демонстрирует публикация статей [7,8]. К тому же анализ критических замечаний автора [7,8] в адрес работ [3-6] показывает, что и с точки зрения математики они неправильны, начиная с первого шага рассуждений (рецензия с обсуждением математической ошибки представлена в редакцию АЭЭ). Считаю, что редакция АЭЭ могла бы признать ошибочность работ [7,8] с публикацией опровержения. Критика [7,8] не означает поддержку работ [3-6], которые могут быть и неверны [1], но не по причинам, изложенным в [7,8].

Соотношение (2) если и могло бы быть предметом научного спора, то лишь с той точки зрения, что в технической гидромеханике (гидравлике) оно излагается в других терминах: «интеграл Бернулли», иногда даже «вклад давления в потенциальную энергию» [10,С.62,64], – причем технические термины могут диссонировать с общефизическими. Но уж совсем бесспорной представляется ситуация со статьей [11], в которой, помимо прочего, утверждается (и существенно используется), что фотон не имеет зависящей от координат волновой функции. Но волновая функция в координатном представлении у фотона есть, она вводится в любом курсе квантовой электродинамики (КЭД - см (2.18,2.26) из [12]) и используется в КЭД расчетах (см, например, [13.С.7])! Именно такая волновая функция обсуждается и в книге [14,С.19-21], на которую ссылается [11]. Тем самым статья [11] показывает, на мой взгляд, пример, когда рецензирование не обеспечивает уже не то что правильность опубликованной работы, но хотя бы ее правдоподобие.

Благодарю проф. Д.Б.Зотьева и Г.В.Трещалова за обсуждения и Редакцию АЭЭ за приглашение к дискуссии Трещалов-Зотьев-Соколов (ТЗС-1-1).

**Список литературы**

**Приложение (публикуется с разрешения автора)**

## К дискуссии Трещалов-Зотьев-Соколов (ТЗС-1-1)
*Г.В.Трещалов*

*В [1-10]  опубликована серия работ, которые затрагивают вопросы гидродинамики, в частности гидродинамические эффекты, возникающие  при функционировании гидротурбин. Автором некоторых из этих работ являюсь я, Г.В.Трещалов. Позднее в журнале «Альтернативная  Энергетика и Экология» появились две статьи Д. Б. Зотьева [11] и [12], в которых мои работы были названы лженаучными.*

*С такой оценкой я не могу согласиться, тем более что, по моему мнению, работы [11-12] содержат серьёзные ошибки и не соответствуют основам теоретической и технической гидромеханики.*

**Основания для подобных выводов следующие**:

1. Работы [11-12] искажают базовое уравнение гидродинамики – уравнение Бернулли для потоков жидкости. В статьях Д. Б. Зотьева потенциальный член в этом уравнении необоснованно выбран в форме $mgH/2$, а не $mgH$, как указано в любом учебнике ($m$ – расход массы через сечение канала, $g$ – ускорение свободного падения, $H$ –глубина потока). Действительно, при расчёте энергии массы воды квазибесконечного потока, протекающей через определённое сечение, может возникнуть искушение поставить двойку в знаменатель потенциального члена энергии, относя потенциальную  энергию к центру масс этого потока, находящегося на половине высоты H (лишь для русел прямоугольной формы). Однако физические условия задачи требуют учитывать гидростатическое давление, которое увеличивается по глубине потока и вносит свой вклад при вычислении полного потока энергии, что и учтено в уравнении Бернулли. При этом добавленный в поток энергии вклад от работы сил давления оказывается равным вкладу от потенциальной энергии и удваивает его, приводя к общеизвестному выражению: mgH/2+ mgH/2= mgH. От геометрической формы профиля русла в этом случае зависит лишь величина каждого из этих слагаемых, однако, их сумма для русел любых форм будет равна mgH.

Интеграл Бернулли является одним из выражений закона сохранения энергии в гидродинамике, поэтому любая неточность в нём дискредитирует и все энергетические соображения работ [11-12], и основанные на них заключения (более подробно это изложено в [13] ). Более того, необходимо отметить, что этим ошибочным представлениям посвящена значительная часть текста в работах [11-12].

2. Исходя из показанной выше ошибки, в [11-12] неверно вычисляется число Фруда для безнапорных потоков жидкости при критическом состоянии потока. Получаемое при этом значение равно 0,5. Оно занижено в два раза по сравнению с общеизвестным, так как критическое состояние потока, определяющее границу между бурным и спокойным его состояниями, достигается при числе Фруда равном 1 (без учёта коэффициента Кориолиса), что имеет место при минимуме удельной энергии живого сечения. Более того, рассуждения о «сложной зависимости» критической глубины от числа Фруда для русел различной формы [12. С.135] (причём со ссылкой на книгу [14], в которой речь идет о сложной зависимости критической глубины от удельного расхода) дают основания полагать, что автором [11-12] неверно был понят физический смысл числа Фруда, вследствие чего им и была допущена эта ошибка. В связи с этим не вызывает удивления тот факт, что неучёт автором [11-12] малости числа Фруда перед турбиной критикуется в одной из дискуссионных работ как причина, приводящая к математической несостоятельности анализа в [11-12] моих исследований.

3. Таким образом, сделанное в [11-12] заключение о лженаучности моих работ и якобы нарушении в них закона сохранения энергии считаю некомпетентным. Все выводы в моих статьях сделаны именно на основе уравнения неразрывности потока (закона сохранения массы) и закона сохранения энергии - уравнения Бернулли для свободного безнапорного потока жидкости, которое в самих [11-12] понимается, по моему мнению, в корне неверно. В частности, простые выкладки по расчёту КПД (КИЭВ) турбины, отнесенного к полной входной мощности потока, показывают, что он не превышает 60% (но выгодно отличается от типичных для традиционных свободнопоточных турбин значений этого параметра равных 15-20%). Строго говоря, работы, в которых инженерные расчетные формулы основаны на законе сохранения энергии (как мои), могут быть в какой-то мере неверны, но не могут противоречить закону сохранения энергии подобно тому, как инженерный расчет падения напряжения по формуле закона Ома не может противоречить закону Ома.

4. Кроме того, в [11-12] ошибочно утверждается, что физический эффект, который был проанализирован в [4] и [5], не наблюдался экспериментально. При более полном знакомстве с литературой экспериментальное подтверждение обсуждаемых мною либо аналогичных эффектов в движущейся среде может быть найдено в [1] и в последующих публикациях по турбине Ленёва. С учётом того, что предложенная мной схема двухступенчатой турбины пока не реализована и это является целью настоящих исследований, то публикация работ [11-12] препятствует дальнейшим научным исследованиям в этом направлении и возможной реализации этих планов. Здесь следует отметить, что модель турбины, схематически представленная в моих публикациях, есть лишь, как мне кажется, наиболее наглядный прототип для понимания сути процесса, а конструкция, которую предполагается создать, в настоящее время патентуется и не может быть представлена в открытой печати [15].

5. Отдельно следует отметить, что критикуемая в [11-12] моя формула полной мощности турбины (1) выведена безотносительно к конкретной конструкции турбины, что неоднократно подчёркивалось во всех моих материалах, в частности и в [4], и в [5].

$$E = \rho * L * ( H_1^2 * V_1 * g + H_1 * \frac{V_1^3}{2} - \frac{3}{2} * \sqrt[3]{(H_1 * V_1)^5 * g^2} ) \qquad (1)$$

Поэтому безосновательны выводы автора [11-12], [16] о том, что, по его мнению, неверны все мои расчёты вследствие якобы неработоспособности модели турбины, представленной в [5].

Другим автором в [13] такая же формула (2) была выведена способом, отличным от моего. Это в очередной раз доказывает, что формула (1), приведённая в [5], верна. Разница состоит лишь в том, что некоторые слагаемые в аналогичной (1) формуле (6) работы [13] представлены через число Фруда.

$$E + W = \rho * L * V_1 * g * H_1^2 ( 1 + \frac{Fr}{2} - \frac{3}{2} \sqrt[3]{Fr} ) \qquad (2)$$

Эта формула является универсальной для расчёта мощности гидротурбин любого типа по условию оптимизации потока в нижнем бьефе (на выходе турбины) - напорных или свободнопоточных турбин и при любых числах Фруда входящего потока - меньших или больших единицы. Все имеющиеся в гидроэнергетике формулы для расчёта оптимальной выходной мощности любых гидротурбин в конечном счёте могут быть получены из неё. Однако обсуждаемый в моих статьях гидродинамический эффект усиления мощности проявляется только при числе Фруда входного потока меньше единицы (пункт 2).

6. Считая ошибочными [11-12], я вовсе не отрицаю наличие некоторых проблем и незаконченности моих работ (к сожалению, в связи с публикацией работ [11-12] эти исследования были приостановлены), в том числе отмеченных критиками. В частности, в материалах [1-10] пока не отражён весьма интересный аспект, который является целью наших дальнейших исследований. Это детальное изучение роли гидравлического прыжка за турбиной и диссипации энергии в нем. Необходимость этих исследований подчеркнута в работе некоторых участников дискуссии. Особый интерес представляет тот факт, что такой гидропрыжок, хотя уже и рассматривался в наших материалах, но подробно пока не изучен ни теоретически, ни экспериментально. Исследования в этом направлении мне представляются весьма перспективными.

**Вывод.**

В связи с высказанными замечаниями, работы [11-12], по моему мнению, должны быть тщательно проанализированы специалистами с последующей возможной публикацией их опровержения, поскольку эти работы, на мой взгляд, неверны и препятствуют дальнейшим научным исследованиям в этом направлении.

Настоящая статья подготовлена по запросу редакции журнала «Альтернативная Энергетика и Экология» для завершения дискуссии Трещалов-Зотьев-Соколов (ТЗС-1-1) в соответствии с письмом исх. № 1303 -14 от 7 февраля 2014 года.

**Список литературы**